# Time as a Statistical Variable and Intrinsic Decoherence

Rodolfo Bonifacio

*Dipartimento di Fisica, Università degli Studi di Milano,
INFN and INFM, Sezione di Milano, Via Celoria, 16, 20133, Milano, Italy*

**Abstract**. We propose a novel approach to intrinsic decoherence without adding new assumptions to standard Quantum Mechanics. We generalize the Liouville equation just by requiring the dynamical semigroup property of time evolution and dropping the unitarity requirement. With no approximations and specific statistical assumptions we find a generalized Liouville equation which depends on two characteristic time $\tau_1$ and $\tau_2$ and reduces to the usual equation in the limit $\tau_1 = \tau_2 \to 0$. However, for $\tau_1$ and $\tau_2$ arbitrarily small but finite, our equation can be written as a finite difference equation which predicts state reduction to the diagonal form in the energy representation. The rate of decoherence becomes faster at the macroscopic limit as the energy scale of the system increases. In our approach the evolution time appears, a posteriori, as a statistical variable *as if* time evolution would take place randomly at average intervals $\tau_2$, each evolution having a time width $\tau_1$. A generalized Tam Mandelstam inequality is derived. The relation with previous work by Milburn is discussed. The agreement with recent experiments on damped Rabi oscillations is described.

## INTRODUCTION

The existence of coherent superposition of states is the basic reason of many paradoxical aspects of *quantum mechanics*. The evolution from a coherent superposition state to a statistical mixture is called decoherence. This is a central problem for measurement theory and for the classical limit of quantum mechanics at the macroscopic level. The Schroedinger cat, which can be in a superposition of states dead or alive, or a macroscopic particle which can be in a superposition of "here" and "there" are typical example of paradoxes whose interpretation is still controversial. A superposition state gives *non zero off-diagonal elements of the density operator*, which originate *quantum interference* and non-classical *correlation effects*. von Neuman postulated the reduction to the diagonal form as a result of a measurement. However, this reduction remains mysterious since it cannot be described by a unitary Hamiltonian evolution and does not clarify *how* and *when* state reduction take place and what is the underlying dynamical process. Essentially two approaches to decoherence have been proposed: the most widely used (1,2) takes quantum mechanics as it is appealing to dissipation due to the interaction of the system with the environment or equivalently with the measuring apparatus, tracing at the end, over the many degree of freedom of the environment. The decoherence of the system is usually described by Master Equations (ME) which are derived using reasonable statistical assumptions and *specific but arbitrary models to simulate* the *environment* and *its interaction with the system*. From a conceptual view point, it appears rather peculiar that one has to invoke the environment, dissipation or the measuring apparatus to conclude that the cat must be dead or alive and a big particle must be here or there and not in a superposition of these possibilities with interference between them. It is like to say that the particle is going to be localized here or there because there is friction, and that the macroscopic limit consist just in the fact that the damping becomes stronger in this limit.

A minority of scientists look for a modification of *standard quantum mechanics* so that decoherence becomes *intrinsic*, i.e., not related to any specific model of interaction or entanglement with the universe. The most widely known solution has been proposed in ref.(3). The authors modify phenomenologically the Schroedinger equation adding nonlinear terms to the Hamiltonian so that the system, at Poisson distributed times, undergoes a sudden localization. Their model contains two unspecified parameters: the frequency and the spatial extent of localization. Other similar models have been proposed (4,5). However, a common feature of these models is that *the energy of the system is not preserved*. This looks to us a very peculiar aspect for an intrinsic theory of decoherence, which in our opinion should act on the coherence and not on the energy of the system.

A scheme for intrinsic decoherence has been proposed by the author in 1983 (6) assuming a finite difference Liouville equation with time step $\tau$ (cronon) which has been connected to the time energy uncertainty relation. In ref. (7) it has been shown that this finite difference equation is equivalent to a semigroup master equation of the Lindblad form.

More recently Milburn (8) has proposed a modification of the Liouville equation assuming that: "the system does not evolve continuously under unitary time evolution, but rather on a stochastic sequence of identical unitary transformations" according to a Poisson distribution. However, in Milburn theory time is a parameter, as in standard quantum mechanics and the Poisson distribution is assumed.

In this paper we generalize the Liouville equation without any specific statistical assumption and without using any specific model for the interaction with the environment. We require the semigroup property of time evolution, dropping the unitarity requirement, we obtain an expression for the density operator which, a posteriori, can be interpreted *as if* the evolution time is not a fixed parameter but a *statistical variable* whose distribution is that of the waiting time of "line theory", i.e., a $\Gamma$-Poisson distribution. Our time distribution function contains two characteristic time, $\tau_1$ and $\tau_2$, which appear naturally in the theory as scaling times. They look the analog in time of GRW localization in space. Precisely time evolution appears on as a random process in which $\tau_2$ is the average time step between two evolution and $\tau_1$ is the time width of each evolution. In terms of waiting time statistics, it is like to say that $\tau_2$ is the average interval between the arrival of two "clients" at the teller in a bank and $\tau_1$ is the time each "client" spends at the teller. Therefore, $\tau_1$ and $\tau_2$ have very different physical meaning and in general $\tau_2 \geq \tau_1$. Our time evolution law, for $\tau_1 = \tau_2 \to 0$ gives back the Liouville equation, whereas for $\tau_1$ and $\tau_2$ arbitrarily small but finite give irreversible state reduction to the diagonal form in the energy representation. Assuming $\tau_1 = \tau_2 = \tau$ sufficiently small, we obtain the Milburn ME (8). Unlike in ref. (3) in our treatment energy is a constant of motion. Our evolution equation can be written in the form of a finite difference equation with time step $\tau_2$. Therefore, according to our formalism, one cannot give a continuos and instantaneous description of time evolution but only at time steps given by the cronon $\tau_2$. From the finite difference equation we obtain a generalized Tam Mandelstam inequality which connects $\tau_1$ and $\tau_2$ with the time-energy uncertainty relation. We apply our formalism to many examples which can be experimentally tested. In particular we derive an extra diffusion term in the position spread of a free particle, decoherence of a free particle prepared in a Schroedinger cat-like state of two different positions; we demonstrate the existence of an intrinsic linewidth for a single mode e.m. field. Finally, we describe decoherence for a spin superposition state in a magnetic field, cancellation of EPR correlation under proper conditions, intrinsic damping of Rabi oscillation in a two level system, in agreement with recent experimental results (9, 10), is predicted.

## THE FINITE DIFFERENCE LIOUVILLE EQUATION

The unitary time evolution of a quantum system is generally described by the Liouville-von Neumann equation, $\partial \rho / \partial t = -iL\rho(t)$, where $\rho$ is the density operator, L is the Liouvillian super-operator, $L\rho \equiv (1/\hbar)[H,\rho]$, and H is the Hamiltonian. Its formal solution can be written as $\rho(t) = e^{-iLt}\rho_0 = e^{-\frac{i}{\hbar}Ht}\rho_0 e^{\frac{i}{\hbar}Ht}$. Here the time t appears as a parameter, not an "observable" or a statistical variable as, for example, are position, momentum and H. In the energy basis, H|n>=E|n>, and one has:

$$\dot{\rho}_{n,m} = -i\omega_{n,m}\rho_{n,m} \qquad \text{and} \qquad \rho_{n,m}(t) = \rho_{n,m}(0)e^{-i\omega_{n,m}t} \qquad (1)$$

where $\omega_{n,m} = (E_n - E_m)/\hbar$. The degenerate case can be included in this notation by assuming that the same states |n> belong to the same eigenstate $E_n$. Let us define a generalized density operator $\bar{\rho}$ defined as a coarse grain time average of $\rho(t)$:

$$\bar{\rho}(t) = \int_0^\infty dt' P(t,t')\rho(t') ; \qquad t,t' \geq 0 \qquad (2)$$

where P(t,t') is a function still to be determined. In particular if P(t,t') = $\delta$(t-t'), $\bar{\rho}(t) = \rho(t)$, so that $\bar{\rho}(t)$ is just a generalized form of $\rho(t)$, because P is unspecified. Equation (2) can be written as

$$\bar{\rho}(t) = \overline{V}(L,t)\rho(0) ; \qquad \overline{V}(L,t) = \int_0^\infty dt' P(t,t')e^{-iLt'} . \qquad (3)$$

We now determine P(t, t') imposing the following conditions:
i) $\bar{\rho}(t) = \bar{\rho}^+(t) \geq 0$; $\text{Tr}\bar{\rho}(t) = 1$ and ii) $\overline{V}(t+t') = \overline{V}(t)\overline{V}(t')$.
Condition (i) identifies $\bar{\rho}(t)$ as a density operator. Condition (ii) is the so called semigroup property which *ensures translational invariance* of the initial condition i.e., $\bar{\rho}(t+t') = \overline{V}(t+t')\rho(0) = \overline{V}(t)\bar{\rho}(t') = \overline{V}(t')\bar{\rho}(t)$. Note that (i) and (ii) are satisfied by the usual Liouville operator $V(t) = e^{-iLt}$. We just drop the request of Unitarity $V^+ = V^{-1}$. Condition (i), using Eq.(2) and taking the trace on both sides, leads to:

$$\int_0^\infty dt' P(t,t') = 1; \qquad P(t,t') \geq 0 \qquad (4)$$

where we used $\text{Tr}\rho(t') = 1$. Equation (4) defines $P(t,t')$ as a probability distribution function; $P(t,t')dt'$ can be read as the probability that the *random variable t* takes a value between t' and t'+dt'. Therefore, $\bar{\rho}(t)$ and $\bar{V}(t)$ appear as an average value respectively of $\rho(t)$ and $V(t)$. The semigroup property (ii) can be generally satisfied assuming

$$\bar{V}(L,t) = [V_1(L)]^{t/\tau_2} \qquad (5)$$

where $\tau_2$ is a scaling time. Note that also the Liouville operator $V(t) = e^{-iLt}$ can be written in the form of Eq.(5) taking $V_1 = e^{iL\tau_2}$ (unitary). We now determine the most general form of $V_1(L)$ compatible with the previous requirements. Let us use the $\Gamma$ function integral identity,

$$(A + iB)^{-k} = \int_0^\infty d\lambda \frac{\lambda^{k-1}}{\Gamma(k)} e^{-A\lambda} e^{-iB\lambda} ; \qquad (6)$$

which is valid for $A > 0$ and $k > 0$. Let us identify the following:
$$V_1^{-1} = A + iB; \qquad k = t/\tau_2; \qquad \lambda = t'/\tau_1$$

where $\tau_1$ is a scaling time generally different from $\tau_2$. In this way by imposing the consistency of Eq.(6) with Eq.(3) for all L and the normalization condition of Eq. (4), we obtain $B = L\tau_1$ and $A = 1$. Therefore, from Eq.(3) and Eq.(5), we have:

$$\bar{\rho}(t) = \bar{V}(t)\rho(0) = \frac{1}{(1+iL\tau_1)^{t/\tau_2}} \rho(0) ; \quad \dot{\bar{\rho}}(t) = -\frac{1}{\tau_2} \ln(1+iL\tau_1)\bar{\rho}(t). \qquad (7)$$

P(t, t') is the $\Gamma$ distribution function:

$$P(t,t') = \frac{1}{\tau_1} \frac{e^{-t'/\tau_1}}{\Gamma(t/\tau_2)} \left(\frac{t'}{\tau_1}\right)^{(t/\tau_2)-1} ; \qquad (t,t' > 0) \qquad (8)$$

This argument can be supported by considering P(t, t')=0 for t'<0 and using the uniqueness of the Fourier transform. Equation (7) provides a generalized form of the density operator and of the time evolution law which has been obtained by requiring the density operator properties (i), the semigroup property (ii) and dropping the unitary requirement. Therefore, we are not adding new elements to the formalism of quantum mechanics but, on the contrary, we are reducing the basic assumptions. Taking $\tau_1 = \tau_2 \to 0$ in Eq.(7) one obtains the Liouville limit and $P(t,t') = \delta(t-t')$. However, as will be shown later, for $\tau_1$ and $\tau_2$ arbitrarily small but finite, one irreversibly approaches the diagonal form in the energy representation.

Equation (7) in the Liouville limit $\tau_1 = \tau_2 \to 0$ recovers again the Liouville equation. When $\tau_1$ and $\tau_2$ are finite, the second order expansion of Eq.(7) gives

$$\dot{\bar{\rho}} = -i\frac{L\tau_1}{\tau_2}\bar{\rho} - \frac{\tau_1^2}{2\tau_2} L^2 \bar{\rho} \qquad (9)$$

where $L^2\rho = [H,[H,\rho]]$. Equation (9), with $\tau_1 = \tau_2$, gives the well known "phase-destroying" ME, deduced by many authors using a reservoir interaction model (2, 11, 12) or specific statistical assumptions (8), and used in Quantum Non Demolition (QND) measurement theory (11). It is straightforward to show from Eq.(7) that $\bar{\rho}(t)$ obeys the following finite difference equation

$$\frac{\bar{\rho}(t) - \bar{\rho}(t - \tau_2)}{\tau_1} = -iL\bar{\rho}(t) = -\frac{i}{\hbar}[H, \bar{\rho}(t)]. \tag{10}$$

Equation (10), for $\tau_1 = \tau_2 \to 0$, again gives the Liouville equation, whereas for $\tau_1 = \tau_2 = \tau$, it reduces to an equation proposed a long time ago (6) to describe irreversible state reduction to the diagonal form. The difference here is that now it has been derived under very general assumptions and two characteristic times appear, with very different physical meanings. In ref. (7), it has been shown that for $\tau_1 = \tau_2 = \tau$, Eq.(10) is equivalent to a semigroup ME of the Lindblad form. The same considerations apply here to Eq.(10), taking $\tau_2 = \tau$ and substituting H with $\bar{H} = H\tau_1/\tau_2$. In this way Eq.(10) is formally identical to that considered in ref. (6).

Equation (10) shows a very important feature. If $\bar{\rho}(t)$ is a solution of Eq.(10), then $f(t)\bar{\rho}(t)$ is also a solution, provided $f(t+\tau_2) = f(t)$. Therefore, $\bar{\rho}(t)$ is uniquely determined only within the "cronon" $\tau_2$, i.e. for time intervals $t = k\tau_2$, with k integer. This fact implies a redefinition of ρ(0) which, in the standard description, is the density operator determined at some *instant* t=0. This is clearly an artifact, because the possibility of an *instantaneous measurement* of a complete set of observables to determine ρ(0) at the instant t = 0 appears as a mathematical abstraction. Our finite interval description of time evolution, which follows from Eq.(10), appears much more realistic because ρ(0) can be interpreted as the density operator determined in a *cronon* interval $\tau_2$. The evolution of Eq.(10) on the "time grid" gives ρ at later time intervals, k=t/$\tau_2$ ≥ 1, and can be parametrized as $\bar{\rho}(k) - \bar{\rho}(k-1) = iL\tau_1\bar{\rho}(k)$, with $k \geq 1$, or equivalently $\bar{\rho}(k+1) = (1 + iL\tau_1)^{-1}\bar{\rho}(k) \approx e^{-iL\tau_1}\bar{\rho}(k)$, for $\tau_1$ small enough. It is *as if* time evolution occurs discontinuously in "quantum jumps" (12) spaced by intervals $\tau_2$ with each evolution taking place within a width $\tau_1$. Each "jump" can be approximated by a unitary time evolution only if $\tau_1$ is small enough. Accordingly, Eq.(3) and Eq.(7), for $t/\tau_2 = k \geq 1$, with $\Gamma(k) = (k-1)!$, can be written as:

$$\bar{V}(k) = \int_0^\infty dt' P(k,t') e^{-iLt'} \; ; \qquad P(k,t') = \frac{1}{\tau_1}\frac{e^{-t'/\tau_1}(t'/\tau_1)^{k-1}}{(k-1)!}, \tag{11}$$

where P(k, t') can be interpreted in two ways: either i) as the well known Poisson distribution in k or ii) as the Γ-distribution function in the continuous variable t'. Unlike previous treatments (8) we adopt the latter. Equation (11) can be interpreted in terms of the *waiting time statistics* for k independent events. According to Eq.(11), *time evolution is made up of random "events"*, i.e., unitary time "evolution". The probability density for k=t/$\tau_2$ events to take place by a time t' is given by Eq.(11). In particular, $\tau_2$ is the *average interval* between two "events" ($\tau_2^{-1}$ is the rate) and $\tau_1$ is the *time width* of each event. Therefore , the "effective evolution time is given by $\bar{t} = (t/\tau_2)\tau_1$. We propose an interpretation of our formalism in terms of a *continuos measurement theory*: $\tau_2^{-1}$ is the *observation rate*, $\tau_1$ is the *time width* of each observation. As we shall see, this interpretation has a precise meaning in laser-micromaser theory of ref. (13). In our formalism the interaction with the environment or with the measurement apparatus is described by two characteristic time $\tau_1$ and $\tau_2$: this characterization is *intrinsic* and *model* *independent* in the sense that it does not depend on the way the measurement is carried out or on the detail on the measurement apparatus. Because the Hamiltonian is a constant of motion, our formalism can be applied, though not exclusively, to a QND measurement. According to this interpretation one obtains a dynamical description of von Neumann state reduction. In fact, in general, in the energy representation Eq.(7) becomes

$$\overline{\rho}_{n,m}(t) = \frac{1}{\left(1+i\omega_{n,m}\tau_1\right)^{t/\tau_2}} \rho_{n,m}(0) = \frac{e^{-i\nu_{n,m}t}}{\left(\sqrt{1+\omega_{n,m}^2\tau_1^2}\right)^{t/\tau_2}} = e^{-\gamma_{n,m}t}e^{-i\nu_{n,m}t}\rho_{n,m}(0) \quad (12)$$

where $\gamma_{n,m} = (1/2\tau_2)\ln(1+\omega_{n,m}^2\tau_1^2)$ and $\nu_{n,m} = (1/\tau_2)\text{arctg}\,\omega_{n,m}\tau_1$, which describes an irreversible evolution. Note that irreversibility is obtained also in the limit $\tau_1 \to 0$ and $\tau_2 \to 0$, provided that $\tau_1^2/\tau_2$ is finite. In general, the decoherence rates $\gamma_{n,m}$ increase as $\tau_2$ becomes smaller or $\tau_1$ becomes larger. Assuming, for simplicity, non degeneracy, for $E_n = E_m$, therefore, $\omega_{n,m}=0$, and $\overline{\rho}_{n,n} = \rho_{n,n}(0)$, so that the energy is a constant of motion, whereas for $n \neq m$, $|\overline{\rho}_{n,m}| \to 0$ with a rate $\gamma_{n,m}$. Therefore, $\overline{\rho}(t) \to \sum_n \rho_{n,n}(0)|n\rangle\langle n|$ i.e., $\overline{\rho}$ approaches the stationary diagonal form. Then a pure state remains a pure state if and only if it is a stationary state; otherwise the system will evolve to a statistical mixture. Note that, at the macroscopic limit, when the energy scale becomes very large, the decoherence time $1/\gamma$ becomes very short. If $\omega_{n,m}\tau_1 << 1$ one has $\gamma_{n,m} \approx \omega_{n,m}^2\tau_1^2/2\tau_2$ and $\nu_{n,m} \approx \tau_1\omega_{n,m}/\tau_2$. This is the result one obtains by the phase diffusing ME, Eq.(9). Therefore, this ME is valid only if $\omega_{n,m}\tau_1 << 1$, for all n and m. This assumption can be made only if the spectrum is bounded so that it cannot be applied to a simple harmonic oscillator, as is usually assumed (10). We emphasize that the basic point of our treatment is that *time appears as a statistical variable* with a distribution function P(t,t'), given by Eq.(8), which for $k=t/\tau_2 = 1$, is a simple exponential. However, for $t/\tau_2 >> 1$, P(t,t') is a strongly peaked function with a mean value, <t'>, and dispersion, σ, given by

$$<t'> = \overline{t} \equiv (t/\tau_2)\tau_1; \quad \sigma = \sqrt{<t'^2> - <t'>^2} = \tau_1\sqrt{t/\tau_2}. \quad (13)$$

Note that <t'> does not coincide with t but with the effective evolution time $\overline{t}$. The dispersion σ scales as $\sqrt{t}$, like in a diffusion process. According to the previous interpretation, the dispersion of σ appears as the dispersion due to $k=t/\tau_2$ statistically independent "events" times the width of each event $\tau_1$. For k=1, σ assumes its minimum value $\tau_1$. Therefore, $\tau_1$ appears as an "inner time", i.e., the intrinsic minimum uncertainty of the evolution time. The relative dispersion, $\sigma/<t'> = (t/\tau_2)^{-1/2}$, goes to zero as the number of evolution steps, $t/\tau_2$, goes to infinity. Furthermore, for $t/\tau_2 >> 1$, P(t,t') can be approximated by a Gaussian in t' with mean value and dispersion given by Eq.(13).

## COMPARISON WITH PREVIOUS WORK

The difference between our treatment and previous descriptions of intrinsic decoherence (8) can be summarized as follows: In ref. (8) it is assumed that i) the system evolves under a random sequences of identical unitary transformations, ii) the probability of n transformations in a time t is given by a Poisson distribution so that (using our notation) $\bar{\rho}(t) = \sum_{n=0}^{\infty} p(n,t) e^{-inL\tau_1} \rho(0)$ where $p(n,t) = \left((t/\tau_2)^n / n!\right) e^{-t/\tau_2}$. In ref. (8) $\tau_1 = \tau_2 = \gamma^{-1}$ is interpreted as a "fundamental time of the universe". Summing up the above series one obtains $\bar{\rho}(t) = \exp\left((e^{-iL\tau_1} - 1)t/\tau_2\right)\rho(0)$ so that

$$\dot{\bar{\rho}}(t) = \frac{1}{\tau_2}\left(e^{-iL\tau_1} - 1\right)\bar{\rho}(t) = \frac{1}{\tau_2}\left(e^{-iH\tau_1/\hbar}\bar{\rho}(t)e^{iH\tau_1/\hbar} - \bar{\rho}(t)\right). \tag{14}$$

The same philosophy has been adopted in laser-micromaser theory (13) substituting $e^{-iL\tau_1}$ with the non unitary operator M (13). In this case, $\tau_1$ is the interaction time, $\tau_2^{-1}$ is the atomic injection rate and Eq. (14) becomes the Scully-Lamb ME. Equation (14), with $\tau_1 = \tau_2 = \gamma^{-1}$, is the basic result of ref. (8). If one applies Eq.(14) with this assumption to calculate the average value of the annihilation operator of an harmonic oscillator, one obtains easily $<\dot{a}> = \gamma(e^{-i\omega/\gamma} - 1)<a>$. From this equation Milburn (8) infers that there are particular frequencies $\omega = 2n\pi\gamma$ where the oscillator is "frozen", i.e., $<\dot{a}> = 0$. As we shall see this unusual behavior does not occur in our treatment because we always have a damped amplitude for all frequencies.

The basic differences between our approach and Milburn's approach (8) are:
i) we do not make any specific statistical assumption: randomness in time evolution appears naturally as an interpretation of our results, i.e., of Eq.(3) and Eq.(8). In particular the Poisson-Γ distribution of Eq.(8) is not assumed but derived.
ii) the choice $\tau_1 = \tau_2$ of ref. (8) implies a severe restriction on the statistical interpretation of the theory.
iii) ref. (8) assumes that the number n of evolution transformations in a time t is a random variable and t is just a parameter as in the Schroedinger equation or in the Liouville equation. In our theory *the waiting time* t for n evolutions to occur is the random variable.

The two view points appear similar but are basically different. In fact, they lead to different results, as one can see by comparing Eq.(14) and Eq.(7). The first can be derived from the second as follows. Using the integral identity $\ln(1+ix) = \int_0^{\infty} d\lambda \frac{e^{-\lambda}}{\lambda}(1 - e^{-i\lambda x})$, and taking $x = L\tau_1$ and $\lambda = t'/\tau_1$, Eq.(7) can be written in the integral form

$$\dot{\bar{\rho}}(t) = \frac{\tau_1}{\tau_2} \int_0^{\infty} dt' \left(\frac{1}{\tau_1} e^{-t'/\tau_1}\right) \left[\frac{e^{-iLt'} - 1}{t'}\right] \bar{\rho}(t) . \tag{15}$$

This equation is completely equivalent to Eq.(7). It can be easily shown that, if the term in square brackets is slowly varying on a time scale $\tau_1$, it can be taken out of the integral with $t' = \tau_1$. Because the term in the round brackets is normalized, one obtains the Milburn Master Eq.(14). Therefore, Eq.(14) can be obtained as an approximation of Eq.(7) or Eq.(15), if $\tau_1 = \tau_2 = \gamma^{-1}$ *is small enough*. In the case of an harmonic oscillator this implies the assumption $\omega/\gamma \ll 1$, which is inconsistent with the condition for freezing $\omega/\gamma = 2n\pi$.

## CLASSICAL LIMIT AND THE GENERALIZED TAM MADELSTAM RELATION

From the finite difference equation (10) and defining the mean value of an observable A in the usual way, $\overline{A} = \text{Tr}\overline{\rho}A$, we obtain

$$\frac{\overline{A}(t) - \overline{A}(t-\tau_2)}{\tau_1} = -\frac{i}{\hbar}\overline{[A,H]}. \qquad (16)$$

Let us apply Eq.(16) to the one dimensional motion of a particle with Hamiltonian $H = p^2/(2m) + V(x)$. Taking A=x and A=$p_x$ we have $(\overline{x}(t) - \overline{x}(t-\tau_2))/\tau_1 = \overline{p}_x(t)$ and $(\overline{p}_x(t) - \overline{p}_x(t-\tau_2))/\tau_1 = \overline{F(x)}$, where $F(x) = -dV/dx$. This is the finite difference version of the Herenfest theorem. The usual form is regained in the continuos limit $\tau_1 = \tau_2 \to 0$, whereas the classical limit is $\overline{F(x)} \approx F(\overline{x})$. Note that the two limits are independent. Therefore, taking only the classical limit one obtains *finite difference Hamiltonian equations* for $\overline{x}$ and $\overline{p}_x$.

We now derive a generalized Tam Mandelstam relation. Using the general uncertainty relation for A and $\overline{H}$ we can write $\sigma(A)\sigma(H) \geq (1/2)\left|\overline{[A,H]}\right| = \hbar\left|\Delta\overline{A}\right|/2\tau_1$, where $\Delta\overline{A} = \overline{A}(t) - \overline{A}(t-\tau_2)$ and we have used Eq.(16). The uncertainty relation can be written in the form $\tau_A\sigma(H) \geq \hbar/2 ; \tau_A \equiv \sigma(A)/\left(\left|\Delta\overline{A}\right|/\tau_1\right)$. This equation appears as a generalized TM inequality and reduces to the usual form in the limit $\tau_1=\tau_2\to 0$. In fact, in this limit one obtains $\tau_A = \sigma(A)/\left|\dot{\overline{A}}\right|$. However, the uncertainty relation can be written as:

$$\frac{\left|\Delta\overline{A}\right|}{\sigma(A)} \leq \frac{\tau_1}{\tau_E} \qquad (17)$$

where $\tau_E = \hbar/2\sigma(H)$ is the "intrinsic inner time" (14) of the system. Because inequality (17) is valid for any observable, we can conclude that if $\tau_1 \leq \tau_E$, then $\left|\Delta\overline{A}\right| \leq \sigma(A)$ necessarily follows. Therefore, no appreciable variation occurs for any observable in the cronon time $\tau_2$. We can say that if $\tau_1 \leq \tau_E$ one has a quasi continuous evolution of the system even using the discrete time description. On the other hand, Eq. (17) states that if $\tau_2$ is such that $\left|\Delta\overline{A}\right| \geq \sigma(A)$, it follows that one must have $\tau_1 \geq \tau_E$. Therefore, the ratio $\tau_1/\tau_E$ rules the rate of change of the state of the system within the time interval $\tau_2$ between two evolutions. The choice $\tau_1 = \tau_E = \hbar/2\sigma(H)$ corresponds to the maximum possible value of $\tau_1$ which guarantees the quasi continuous evolution which is commonly observed. This choice is clearly meaningless in case of a non Hamiltonian system.

8...

## CALCULATION OF PHYSICAL QUANTITIES

Equation (2) allows one to calculate all physical quantities by just performing a time integral of the usual expressions. Let us give some examples. The mean value $\overline{A}(t) = \text{Tr}\overline{\rho}(t)A$ can be easily obtained using by Eq.(2) as

$$\overline{A}(t) = \int_0^\infty dt' P(t,t') <A>_{t'} \; ; \qquad <A>_t = \text{Tr}\rho(t)A . \tag{18}$$

Therefore, a constant of motion remains a constant of motion, whereas oscillating quantities, as $e^{i\omega t}$ will be damped with a rate constant $\gamma = (1/2\tau_2)\ln(1+\omega^2\tau_1^2)$ similarly to $\rho_{n,m}$ as given by Eq.(12). Furthermore, from Eq.(2),

$$\langle x|\overline{\rho}(t)|x'\rangle = \int_0^\infty dt' P(t,t') \langle x|\rho(t')|x'\rangle . \tag{19}$$

The same relation holds for matrix elements in any representation. However, Eq.(19) is of particular relevance. In fact, for x = x' it gives the position probability density, $\overline{P}(x,t)$. In particular, if the initial state is a pure state, $|\psi_0\rangle\langle\psi_0|$, then $\langle x|\rho(t')|x\rangle = |\psi(x,t')|^2$. Furthermore, because the Wigner function is related to the Fourier transform of $\langle x|\rho(t)|x'\rangle$, it can be obtained using the same integral relation as in Eq.(19). It can be easily shown that the same relation is valid for the Glauber P-function. We now make some testable applications of our formalism.

## HARMONIC OSCILLATOR

In the case of the harmonic oscillator the photon number is a constant of motion. On the contrary the amplitude <a> goes to zero. In fact, because $<a>_t = <a>_0 e^{-i\omega t}$, using Eq.(18) and Eq.(6) one obtains $\overline{a}(t) = \left(1/(1+i\omega\tau_1)^{t/\tau_2}\right)<a>_0 = <a>_0 e^{-\gamma t}e^{-i\nu t}$, where $\gamma = (1/2\tau_2)\ln(1+\omega^2\tau_1^2)$ and $\nu = (1/\tau_2)\text{arctg}\,\omega\tau_1$. Therefore, $|\overline{a}(t)|$ goes to zero as t goes to infinity as it must be for a phase-destroying process. The behavior of $\gamma$ and $\nu$ is formally the same as that of $\gamma_{n,m}$ and $\nu_{n,m}$ described before. Therefore, the same discussion applies just abolishing the index n,m. This intrinsic linewidth $\gamma$ should be observable only if it dominates on all other linewidth, i.e., if $\tau_1$ is large enough and $\tau_2$ is small enough. In the opposite case one obtains a lower limit for $\tau_2$ and an upper limit for $\tau_1$.

## FREE PARTICLE LOCALIZATION

Let us consider a free particle prepared in a Schroedinger cat-like state i.e., in a coherent superposition state of two minimum uncertainty wave packets, centered at different positions, $x_1$, and $x_2$. The wave function at time t is given by

$$\psi(x,t) = (1/\sqrt{2})(\psi_1(x,t) + \psi_2(x,t)) \tag{20}$$

where

$$\psi_j(x,t) = \frac{1}{\left(\sqrt{2\pi}\sigma_x\right)^{1/2}} e^{-(x-x_j)^2/4\sigma_x^2} \left(1 - i\frac{\sigma_v}{\sigma_x}t\right) \qquad j=1,2. \qquad (21)$$

Here, for simplicity, we have neglected the Schroedinger spread, by assuming $\sigma_v$ to be small enough. This is certainly the case at the macroscopic limit, because $\sigma_v = \hbar/(2m\sigma_x)$, where m is the mass of the particle. From Eq.(20) we have:

$$P(x,t) = |\psi(x,t)|^2 = (1/2)|\psi_1(x)|^2 + (1/2)|\psi_2(x)|^2 + \text{Int} \qquad (22)$$

where $\text{Int} = |\psi_1(x)||\psi_2(x)|\cos\omega t$, with

$$\omega = 16mE^2 Dx/\hbar^3, \qquad (23)$$

where $x_1 = -x_2 = D/2$ and $E = (1/2)m<v^2>$. Therefore, in standard quantum mechanics the interference term is oscillating in time. In contrast, in our formalism the interference disappears because, using Eq.(18) and Eq.(6), the average of cos(ωt) is exponentially damped at a rate given by $\gamma = (1/2\tau_2)\ln(1+\omega^2\tau_1^2)$, where ω is given by Eq.(23). The interference disappears in a decoherence time $t_D = \gamma^{-1}$. Note that if $\omega\tau_1 \ll 1$ one can expand the logarithm to the first order, obtaining a decoherence rate proportional to $\omega^2$, i.e., to the square of the distance between the centers of the two packets, in agreement with previous treatments. The interaction with the environment and/or with the measuring apparatus appears quite naturally in our formalism via the two characteristic times without specifying any interaction model. Equation (23), together with the expression of γ, clearly shows that at the macroscopic limit, when m and E are very large, or at the classical limit $\hbar \to 0$ the decoherence rate becomes extremely fast.

## INTRINSIC DECOHERENCE OF RABI OSCILLATIONS

The same considerations can be applied to a system of two level atoms injected one at the time in a high Q resonant cavity and prepared in a Rydberg state so that the atomic and cavity decay time and the intrinsic decoherence are very long. This is the experimental situation in ref. (9). If the field is in a n-Fock state and the atoms are injected in the excited state, the atoms will oscillate between the upper and lower state so that the population difference d will oscillate as $d = \cos\Omega t$, where $\Omega = g\sqrt{n+1}$ is the Rabi frequency and g is the one photon Rabi frequency accordingly with the Jaynes-Cummings atom-field interaction Hamiltonian model. In our formalism these oscillations will be damped with a rate constant, $\gamma = (1/2\tau_2)\ln(1+\Omega^2\tau_1^2)$. Therefore, if γt = γL/v >> 1, the population difference will approach the steady state value d=0 so that the population of the upper level will approach the value 0.5. This behavior has been indeed observed in ref.(9) also for the vacuum state where dissipation losses are ineffective and in ref. (10) for a n-Fock state, in a different experimental situation. In particular, in ref.(10), it has been observed an increase of the damping rate γ with n, in "qualitative" agreement with $(n+1)^{0.7}$. Let us point out that our damping rate has a logarithmic dependence on (n+1), which, if $\tau_1$ is small enough, goes like (n+1). Whether this damping is due to experimental problems, as suggested by the authors, or can be attributed to our intrinsic decoherence mechanism deserves further experimental and theoretical investigation. Obviously the same intrinsic damping can be predicted for the Rabi oscillations in a NMR configuration.

# CANCELLATION OF EPR CORRELATION

Let us assume two spin-1/2 neutral particles, say 1 and 2, traveling in opposite directions with velocity v and prepared in the singlet entangled state $|\psi_0\rangle = (1/\sqrt{2})(|+_1,-_2\rangle - |-_1,+_2\rangle)$ with a constant magnetic field $B_0$ in the z direction acting on the path of particle 1 for a length L. We have $|\psi\rangle_t = (1/\sqrt{2})(e^{i\omega_0 t/2}|+,-\rangle - e^{-i\omega_0 t/2}|-,+\rangle)$ where $\omega_0$ is the Larmor frequency. The associated density operator reads

$$\rho(t) = |\psi(t)\rangle\langle\psi(t)| = \left(\rho_D - \frac{1}{2}e^{i\omega_0 t}|+,-\rangle\langle-,+| - h.c\right) \quad (24)$$

where $\rho_D$ is the diagonal part. Note that for $\omega_0 L/v = 2n\pi$ one has again the initial singlet state. Therefore, the presence of a e.m. field would become ineffective regarding EPR correlation under these conditions. However, in our formalism the oscillation of the off-diagonal terms disappears exponentially. Therefore, if $\gamma L/v \gg 1$ the EPR correlation should disappear even if $\omega_0 L/v = 2n\pi$. Any evidence of correlations smaller than the one expected in the conditions described above, would be a strong support of our theory.

# CONCLUSION

We have generalized the Liouville equation without introducing new assumptions to quantum mechanics. On the contrary, we have reduced the basic axioms of quantum mechanics dropping the unitarity condition and maintaining only the semigroup property of the time evolution operator. The new equation describes intrinsic decoherence giving irreversible state reduction to the diagonal form in the energy basis. Several testable examples have been discussed. An intrinsic linewidth for a harmonic oscillator, space localization of a free particle prepared in a Schroedinger cat-like state and disappearance of EPR correlations under proper conditions have been described. In particular, damping of Rabi oscillations, in agreement with recent experimental results (9, 10) has been predicted.

# ACKNOWLEDGEMENT

We must acknowledge that our work started in 1983 stimulated by Caldirola's pioneering work on a finite difference Schroedinger equation to describe the quantum theory of the electron (15). We are indebted to L. Davidovich and N. Zagury for their hospitality at the Federal University of Rio de Janeiro where this work has been partially developed under their continuous stimulation, suggestions and criticisms. We are grateful to S. Osnaghi for his continuous collaboration, L. De Salvo for her assistance and to D. A. Jaroszynski for reading, improving the manuscript and fruitful discussions.